\documentclass[a4paper,11pt]{article}
\pdfoutput=1 % if your are submitting a pdflatex (i.e. if you have
\usepackage[T1]{fontenc} % if needed
\usepackage{graphics}
\usepackage{graphicx}
\usepackage{float}
\usepackage{caption}
\usepackage{subcaption}
\usepackage{bbding}
\usepackage{jheppub} 
\usepackage{color}		
\usepackage{slashed}		
\usepackage{multirow}
\usepackage{adjustbox}
\usepackage{hyperref}

\title{Deviations of exact neutrino textures using radiative neutrino masses}
\author{D. Wegman $^{1}$}
\affiliation{
	$^1$ IFPA, STAR institute, Universit\'e de
	Li\`ege, B\^at B5, Sart Tilman B-4000 Li\`ege 1,
	Belgium} 

\emailAdd{dwegman@uliege.be}
\abstract{
	The Weinberg operator allows for the construction of radiative Majorana neutrino masses. In this letter, it will be shown that it is possible to construct a one-loop diagram that will be the principal component of the neutrino mass matrix and that will have an exact mixing matrix with $\theta_{13} =0$. The addition of a two-loop diagram, which is naturally suppressed, allows the creation of the correct perturbations that will give a neutrino mixing matrix with  entries inside experimental constrains, including the possibility of large CP Dirac phases.}


\begin{document} 
	\maketitle
	
	\section{Introduction}
	
	There are three observations that give evidence that the Standard Model (SM) is incomplete: the baryon asymmetry of the Universe, the cosmological evidence of Dark Matter (DM), and the existence of neutrino masses. While LHC is still trying to find evidence of new physics, neutrino experiments have successfully collected huge amounts of data, making the neutrino sector a promising area to probe into. Specifically six years ago many neutrino experiments, such as T2K \cite{T2K}, MINOS  \cite{MINOS}, RENO \cite{RENO}, Daya Bay \cite{Daya Bay}, and Double Chooz \cite{Double Chooz},  found that the smallest of the neutrino mixing angles is in fact non-zero. This led to the consequence that neutrinos cannot have exact mixing textures such as tribimaximal, bimaximal, etc.  
	Having three non-zero mixing angles has the added effect that Dirac phases cannot be absorbed into the fields, making CP violation in the neutrino sector a viable option. 
	
	The  current neutrino global fit \cite{deSalas:2017kay} shows  as follows:
	
	\begin{table}[h]\centering
		\begin{tabular}{lcc}
			\hline
			Parameter & Best fit $\pm$ $1\sigma$ &   3$\sigma$ range
			\\
			\hline
			$\Delta m^2_{21}\: [10^{-5} ] \text{eV}^2$
			& 7.60$ \pm 0.19$  & 7.05--8.14 \\
			$|\Delta m^2_{31}|\: [10^{-3}]\text{eV}^2$ (NH)
			&  2.55$\pm 0.04$ &    2.43--2.67\\
			$|\Delta m^2_{31}|\: [10^{-3}]\text{eV}^2$ (IH)
			&  2.49$\pm 0.04$ &  2.37--2.61 \\
			$\sin^2\theta_{12} / 10^{-1}$
			& 3.21$^{+0.18}_{-0.16}$  &  2.73--3.79\\
			$\sin^2\theta_{23} / 10^{-1}$ (NH)
			&	4.30$^{+0.20}_{0.18}$ 
			& 3.84 -- 6.35 \\
			$\sin^2\theta_{23} / 10^{-1}$ (IH)
			& 5.96$^{+0.17}_{-0.18}$  & 3.88--6.38 \\
			$\sin^2\theta_{13} / 10^{-2}$ (NH)
			& 2.155$^{+0.090}_{-0.075}$  & 1.89--2.39 \\
			$\sin^2\theta_{13} / 10^{-2}$ (IH)
			& 2.140$^{+0.082}_{-0.085}$ & 1.89--2.39 \\
			$\delta/^{\circ}$  (NH)& 252$^{+56}_{-36}$   &0-360 \\
			$\delta/^{\circ}$ (IH)& 259$^{+47}_{-41}$   &0-360 \\
			\hline
		\end{tabular}
		\caption{ \label{tab:summary} Neutrino oscillation parameters
			summary.}
	\end{table}
	
	As indicated in Table \ref{tab:summary}, among many things, one can see that even at the 3$\sigma$ range,  $\theta_{13}$ (although smaller than the other two mixing angles) has a non null value. Also,  the table shows that the central value for the CP phase is large ($\sim 3\pi/2$), even though at the 3$\sigma$ range the value is still undetermined.

	Still, given the proximity of the mixing angles to exact values, one can think of models that start from an exact texture and make deviations from it \cite{Rodejohann:2011uz,Chu:2011jg,Toorop:2011jn,BhupalDev:2011gi,Chao:2011sp,Ma:2011yi,Branco:2014zza,Bhattacharya:2012rz,He:2011gb,King:2013eh,Ma:2015pma}. A complete model should answer the question of where these deviations come from and  why they are smaller than the original values of the model.
	
	The nature of neutrinos is still not known. As they have no electric charge, they could be different from the SM fermions which are Dirac particles. In fact, neutrinos could be their own antiparticles and therefore be Majorana particles.
	
	The study of Majorana neutrinos can be done using effective theory to write a Lagrangian. First, the SM Lagrangian is used. Secondly, higher order non-normalizable operators are added, 
	\begin{equation}
	\mathcal{L}= \mathcal{L_\text{SM}} +  \sum_{n>4}  \mathcal{L_\text{n}}.
	\label{eq:higer}
	\end{equation}

	It can be shown~\cite{Weinberg:1979sa} that there is only one possibility for $n=5$, commonly referred to in literature as the Weinberg dim=5 operator, 
	\begin{equation}
	\mathcal{L}_5 \sim \frac{c_{\alpha \beta}}{\Lambda} \left(\overline{L^\texttt{c}_{\alpha}} \tilde{H}^*\, \tilde{H}^{\dagger}L_{\beta}\,\right),
	\label{eq:weinberg}
	\end{equation} 
	where $H$ is the Higgs field, $L$ the SM lepton field,  $\Lambda$ is the energy scale, and $c$ is a constant.
	
	In this letter  a method to perturb any exact mixing texture will be proposed. The main contribution to the neutrino masses is a radiative one-loop (1-L) Majorana mass diagonalized by an exact texture. Subsequently, a two-loop (2-L) Majorana mass can be added. This contribution is naturally suppressed and explains the deviations from the exact texture.
	
	If one wishes to construct a 1-L dim=5 Majorana neutrino diagram that has no tree-level contributions,  there are only four realizations \cite{Bonnet:2012kz}. Although, after electro-weak symmetry breaking (EWSB), in their mass eigenstates, all possibilities get reduced to the same one; a  diagram with two neutrinos as external legs and an internal loop with a fermion and a scalar.
	
	The number of 2-L dim=5 Majorana neutrino diagrams increases greatly compared to the 1-L, and although there is no study  looking at all possibilities, some realizations can be seen in ref. \cite{Sierra:2014rxa}. Nevertheless, after EWSB all the 2-L diagrams get reduced to three types: with one, two or three fermion messengers, plus the necessary scalars to close the loops. In the literature one can find multiple examples of one-loop, two-loop, and even higher radiative neutrino models \cite{Ma:2006km,Restrepo:2013aga,Bonnet:2009ej,Babu:2001ex,Hirsch:2004he,Simoes:2017kqb,Sierra:2016rcz,Sierra:2016qfa,Nomura:2017ezy,Cai:2017jrq,Babu:2002uu,Aoki:2017eqn,McDonald:2003zj}.
	
	Given what was just stated, it is easy to realize that all 1-L diagrams from a Weinberg operator will lead to a neutrino mass that  is proportional to $Y Y^T$ (this can be seen in Fig.\ref{fig:exoneloop})  times the loop integral, where $Y$ is a generic $3 \times n$ Yukawa matrix, with $n$ the number of generations for the internal fermions.

	In this paper, a minimal model will be used as an example of the method.  It will be shown that while the 1-L contribution reproduces a mass matrix that will be diagonalized by a tribimaximal mixing, the full mass matrix will agree with experimental results. Including ( In the right portion of the parameter space) large Dirac phases.

	\section{Example}
	
\begin{figure}[h]
		\centering
		\includegraphics[width=4cm,height=2.4cm]{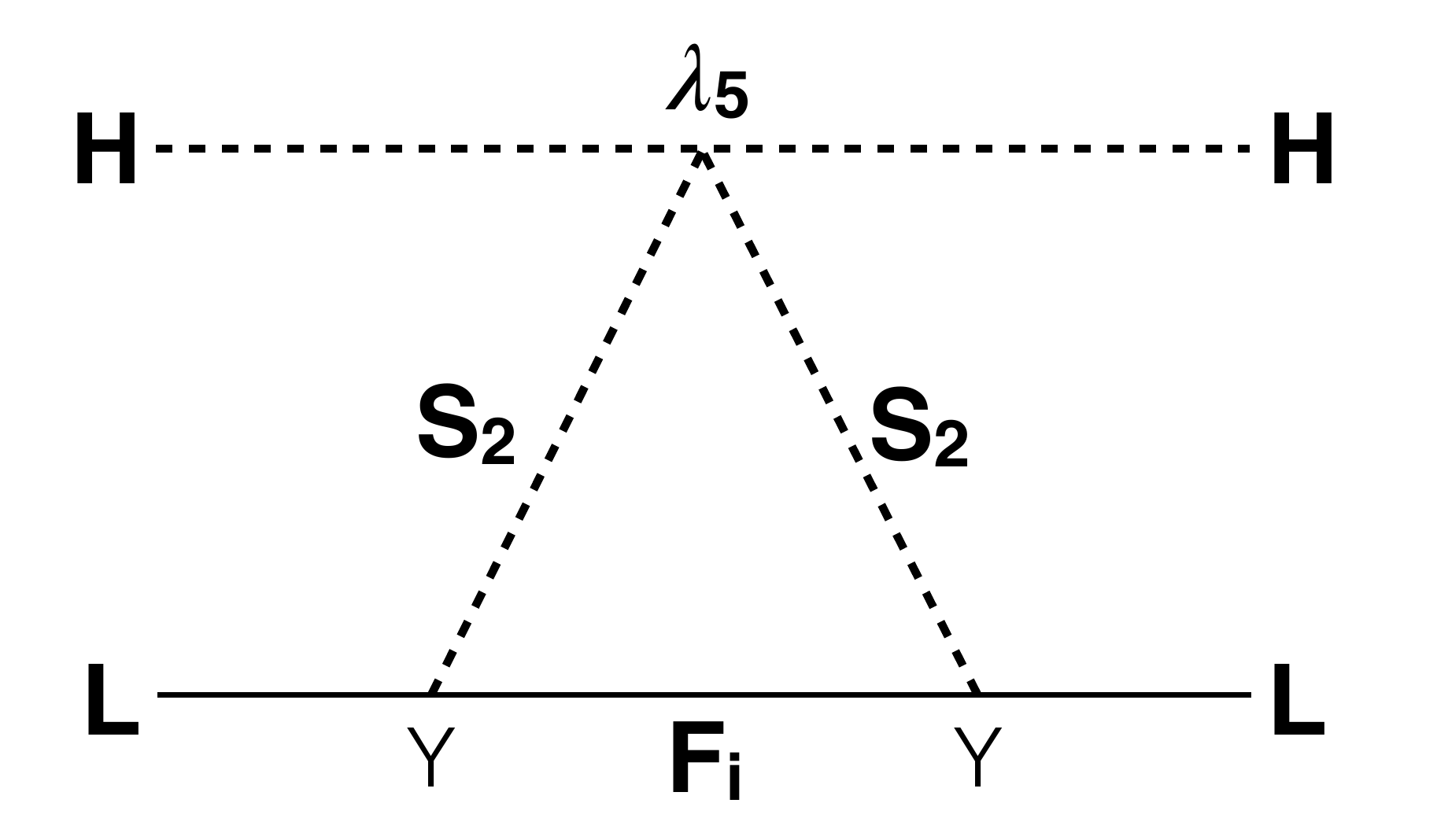}
		\includegraphics[width=4cm,height=2.4cm]{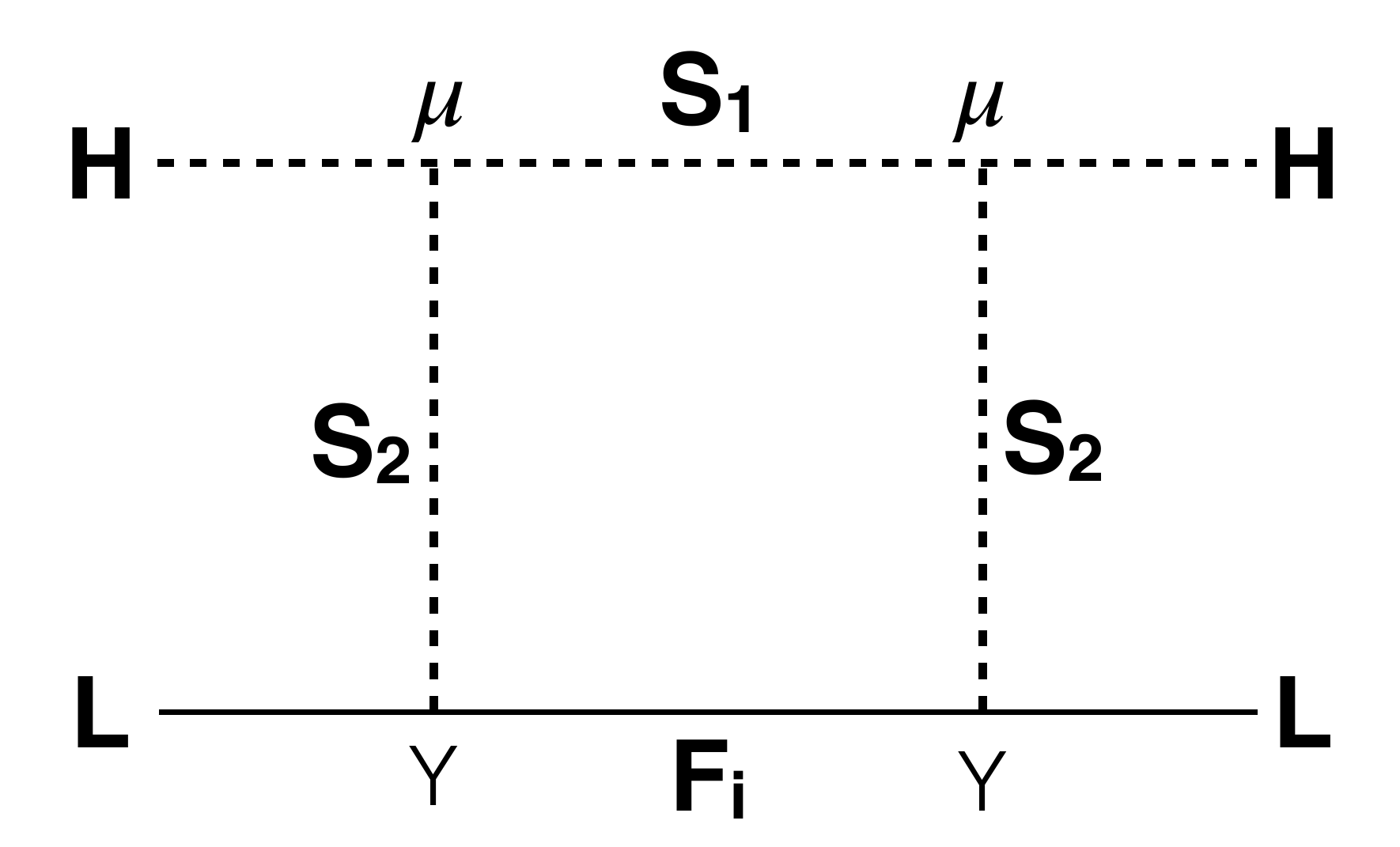}
		\caption{One-loop (1-L) contributions to neutrino mass. }
		\label{fig:exoneloop}
	\end{figure}

\begin{figure}[h]
		\centering
		\includegraphics[width=6cm,height=2.4cm]{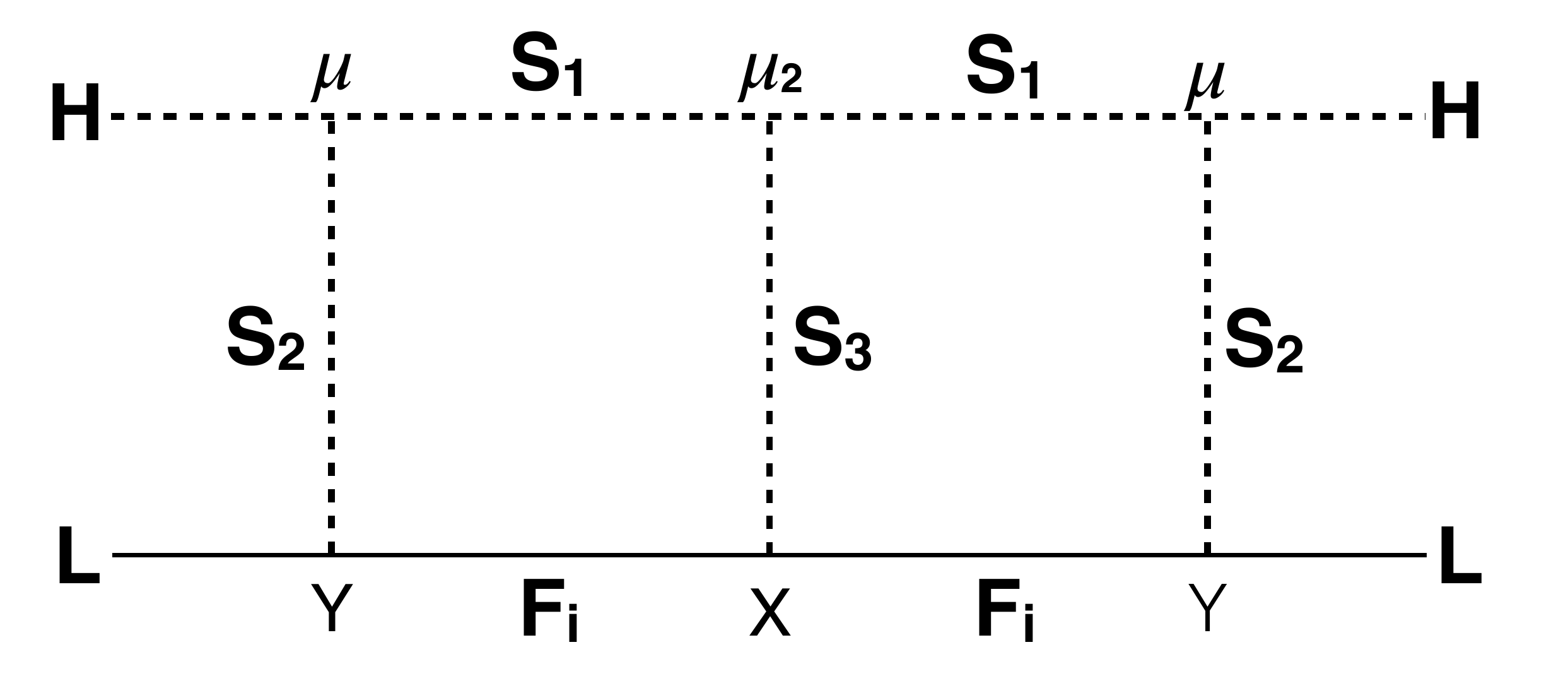}
		\includegraphics[width=6cm,height=2.4cm]{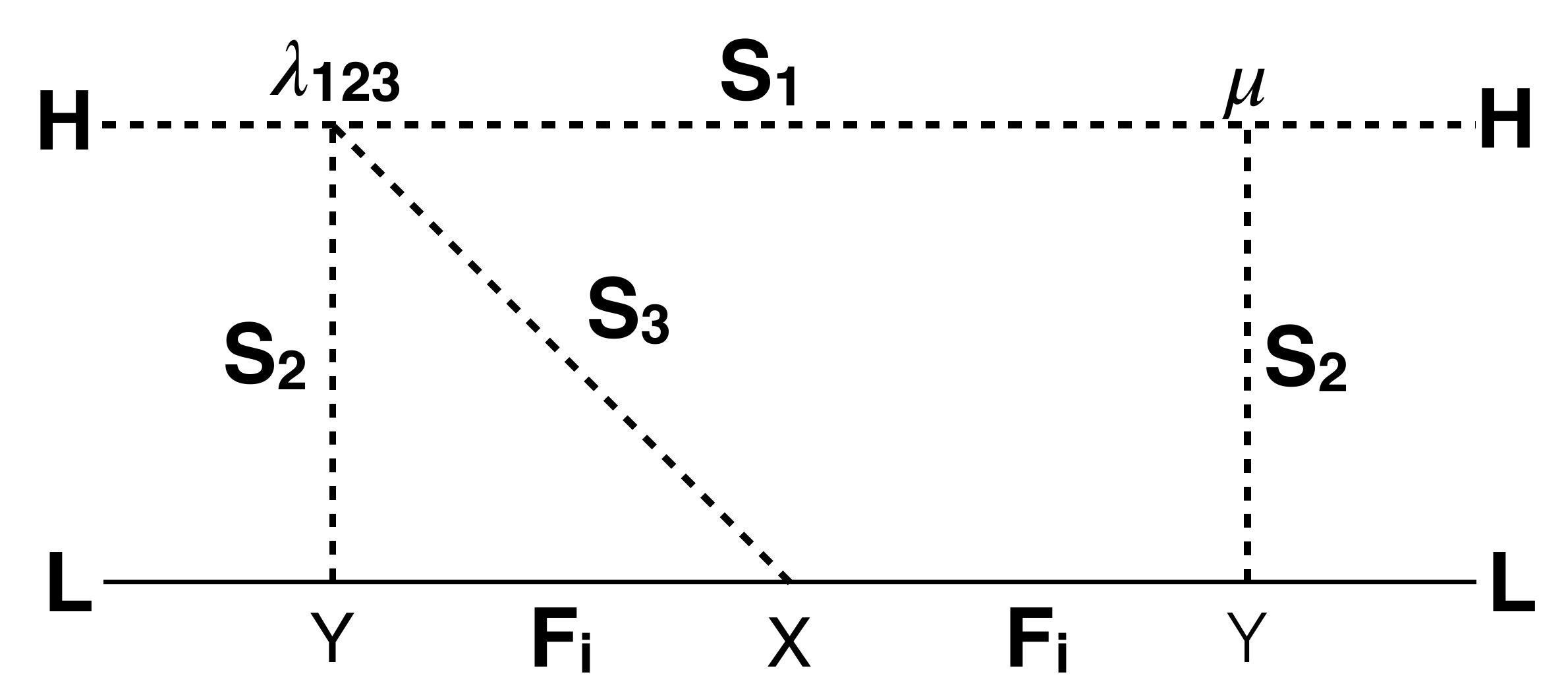}
		\caption{Two-loop (2-L) contributions to neutrino mass. }
		\label{fig:extwoloops}
	\end{figure}

	To demonstrate the method an example will be used. Table \ref{tab:part} shows the particle content, and Figs.$\,$\ref{fig:exoneloop} and \ref{fig:extwoloops} the diagrams for  the 1-L and 2-L contributions to the neutrino mass that can be generated with this particle content (while other diagrams can exist for the 2-L case, they will not affect the structure of the mass matrix, and therefore for simplicity we will assume them to have a null contribution).

	The scalar content for these models is 2 doublets with hypercharge $Y \! \!\!=\!1$, and two singlets with null hypercharge. Therefore, the content is no different than a 2HDMS inert model \cite{Grzadkowski:2009iz,Boucenna:2011hy}, where one of the doublets acts as the SM Higgs, and all other scalars have no vev. The singlets won't mix with the doublets or affect their masses.

	Other than the BSM scalars, three singlet fermions are needed. Although the loop can be closed with only one extra fermion, three are needed to have three non-zero neutrino masses.
	
	\begin{table}[h]\centering
		\begin{tabular}{|c|c|c|c|}
			\hline 
			& $SU(2)_{L}$ & $U(1)_{Y}$ & $Z_{2}$\tabularnewline
			\hline 
			\hline 
			$H$ & 2 & 1 & +\tabularnewline
			\hline 
			$L$  & 2 & -1 & +\tabularnewline
			\hline 
			$S_{1}$ & 1 & 0 & -\tabularnewline
			\hline 
			$S_{2}$ & 2 & 1 & -\tabularnewline
			\hline 
			$S_{3}$ & 1 & 0 & +\tabularnewline
			\hline 
			$F_{i}$ & 1 & 0 & -\tabularnewline
			\hline 
		\end{tabular}
		\caption{\label{tab:part} Particle content.}
	\end{table}

	A  $Z_2$ symmetry needs to be included to avoid FCNC \cite{Haber:2006ue,Ginzburg:2004vp,Branco:2011iw}. To do so, any doublet scalar of the model with the same quantum numbers as the Higgs needs to be odd under this symmetry. Moreover, the symmetry is necessary to forbid tree-level contributions to the neutrino mass, both Majorana and Dirac terms. This means that any fermionic singlet (or triplet) with $Y \! \!\!=\!0$ must, also, be odd under the $Z_2$. Applying these details, and given the interactions needed for the creation of the diagrams in Figs.$\,$\ref{fig:exoneloop} and \ref{fig:extwoloops}, one can set the $Z_2$ of all fields in the model.
	
	In addition, the existence of a $Z_2$ symmetry gives as a consequence a stable neutral particle that can be a DM candidate.
	
	The relevant Lagrangian terms that are used to build the 1-L and 2-L diagrams are
	\begin{equation}
	\begin{aligned}
	\mathcal{L}^{}&= 
	Y^{ia}\,(\overline{L^\texttt{C}}_i\,P_L) {F}_a \,S_2\,+\,
	Z^{ab}\,\overline{F}_a\, {F}_b\, S_3 +\,\text{h.c.}\,,
	\end{aligned}
	\label{eq:L1b}
	\end{equation}
	and the relevant part of the scalar potential is
	\begin{eqnarray}
	\begin{aligned}
	V&=
	\mu\,H\,S^\dagger_1\,S_2
	\,+\,\mu_2\,S_1\,S^\dagger_1\,S^\dagger_3\,  +\, \frac{1}{2}\lambda_5 (H^{\dagger}S_2) \nonumber \\
	&+ \lambda_{123} H^\dagger S_2 S_1^* S_3
	+\,\text{h.c.}\,
	\end{aligned}
	\end{eqnarray}
	
	Note that given the quantum numbers of $S_3$, the Yukawa matrix $Z^{ab}$ is symmetric.
	
	The neutrino mases are:
	\begin{eqnarray}
	m_{1-L}&=&Y \Lambda_1 Y^T,  \label{masses1L}  \\
	m_{2-L}&=& Y  \Lambda_{21} Z \Lambda_{22} Y^T \equiv Y  \Lambda_{2}   Y^T   \label{masses}
	\end{eqnarray}

	The integrals contained in $\Lambda_1$ and $\Lambda_2$ have been explicitly solved in ref.\cite{Sierra:2014rxa}.  Nevertheless, we will use the approximation $M_F<<m_{s3}<<m_{s2}<<m_{s1}$ that will simplify the expression for the neutrino masses at the 1-L and 2-L level \cite{Ma:2006km,Babu:2002uu},

	\begin{eqnarray}
	m_{1-L}&=&\frac{ \; (\lambda_5)_{eff} \; v^2 }{16  \pi^2 \; m_{s2}^2} Y\cdot M_F \cdot Y^T ,\\ 
	m_{2-L}&=&\frac{\mu_{eff}}{(96 \pi^2) \; m_{s2}^2} Y\cdot M_F \cdot Y_2\cdot M_F \cdot Y^T
	\end{eqnarray}
	
	where $M_F$ is a $3 \times 3$ diagonal matrix with the masses of the BSM fermions $F_i$ ($m_{f1},m_{f2},m_{f3})$, $(\lambda_5)_{eff}$ and  $(\mu)_{eff}$ are effective couplings that can be used given that after EWSB both 1-L and both 2-L contributions reduce to a similar diagrams respectively.

	Using Casas-Ibarra parametrization \cite{Casas:2001sr} we can parametrize the Yukawa matrix $Y$ in  Eq. (\ref{masses}) by
	\begin{eqnarray}
	Y&=& U_0 m_d^{1/2} R_1^T \Lambda_1^{-1/2} \label{Casas} \\
	&=&U_{\nu}M_d^{1/2} R_2^T (\Lambda_1+\Lambda_2)^{-1/2},\nonumber
	\end{eqnarray}
	
	where $m_d$ and $M_d$ are the diagonal matrices containing the eigenvalues of $m_{1-L}$ and $M_{\nu}=m_{1-L}+m_{2-L}$ respectively, and $R_i$ are  $3 \times n$ orthonormal matrices, with n set by the dimensions of $\Lambda_1$ and $\Lambda_2$.  Although the matrices $R_i$  are generically arbitrary, they  have a specific shape once a model (and a discrete symmetry) has been chosen to set the structure of $Y$.

	Applying the parametrization as seen in Eq.(\ref{Casas}) and using $U_0=U_{TBM}$ and $R=\mathbb{I}$, it is possible to construct a simple yukama matrix with only three parameters,

	\begin{equation}
	Y
	=
	\left\lgroup
	\begin{matrix}
	\sqrt{\frac{2}{3}}  K_1& \frac{1}{\sqrt{3}} K_2 & 0  \\
	-\frac{1}{\sqrt{6}} K_1 & \frac{1}{\sqrt{3}} K_2& -\frac{1}{\sqrt{2}} K_3  \\
	-\frac{1}{\sqrt{6}} K_1 & \frac{1}{\sqrt{3}} K_2 & \frac{1}{\sqrt{2}} K_3
	\end{matrix}
	\right\rgroup \,
	\label{Yuk}
	\end{equation}
	
	where $K_i=m_{di}/\Lambda_{i1}$, with $i$ indicating the three different eigenvalues of $\Lambda_{1}$.
	
	The choice for $R=\mathbb{I}$ is not unique, but it is the one that reduces the number of parameters to its minimal. Clearly, if one wishes to use a specific model with a discrete symmetry  other choices should be considered. 
	
	To ensure that the mass differences match the experimental results, we diagonalize  the 1-L mass $m_{1-L}$ with a tribimaximal mixing, $U_{TBM}$,  the diagonal masses are:
	\begin{eqnarray}
	m^0_i=\frac{(\lambda_5)_{eff} \: v^2 K_i^2 \; m_{Fi}}{8 \pi^2 \; m_{s2}^2}.
	\end{eqnarray}
	
	With the previous result we can  use $\left(m^0_j\right)^2-\left(m^0_i\right)^2=\Delta_{ji}^2$  as an input parameter (where $\Delta_{ji}$ is the neutrino mass square differences at the 1-L level), to calculate two of the fermion masses as a function of the third mass,
	
	\begin{eqnarray}
	m_{F1}=\pm\frac{\sqrt{(\lambda_5)_{eff}^2 \: v^4 K_2^4 \; m_{F2}^2-64 \pi^4 m_{s2}^4 \Delta_{21}^2}}{\lambda_5 K_1^2 v^2} \label{mf1}\\
	m_{F3}=\pm\frac{\sqrt{(\lambda_5)_{eff}^2 \: v^4 K_2^4 \; m_{F2}^2\pm 64 \pi^4 m_{s2}^4 \Delta_{32}^2}}{\lambda_5 K_3^2 v^2}. \label{mf3}
	\end{eqnarray}
	
	Using Eqs.(\ref{mf1})-(\ref{mf3}) and the matrix in Eq.(\ref{Yuk}), it is possible to have a neutrino mass matrix that at the 1-L level will always give tribimaximal mixing and that has mass squared differences inside experimental constrains. It is important to note that $\Delta_{ji}^2$  should  be close to the experimental values, but not necessarily have the exact central values, since  the results will change when the 2-L is added to the neutrino masses.
	
\begin{figure}[h]
		\centering
		\includegraphics[width=8cm,height=12cm,angle=270,origin=c
		]{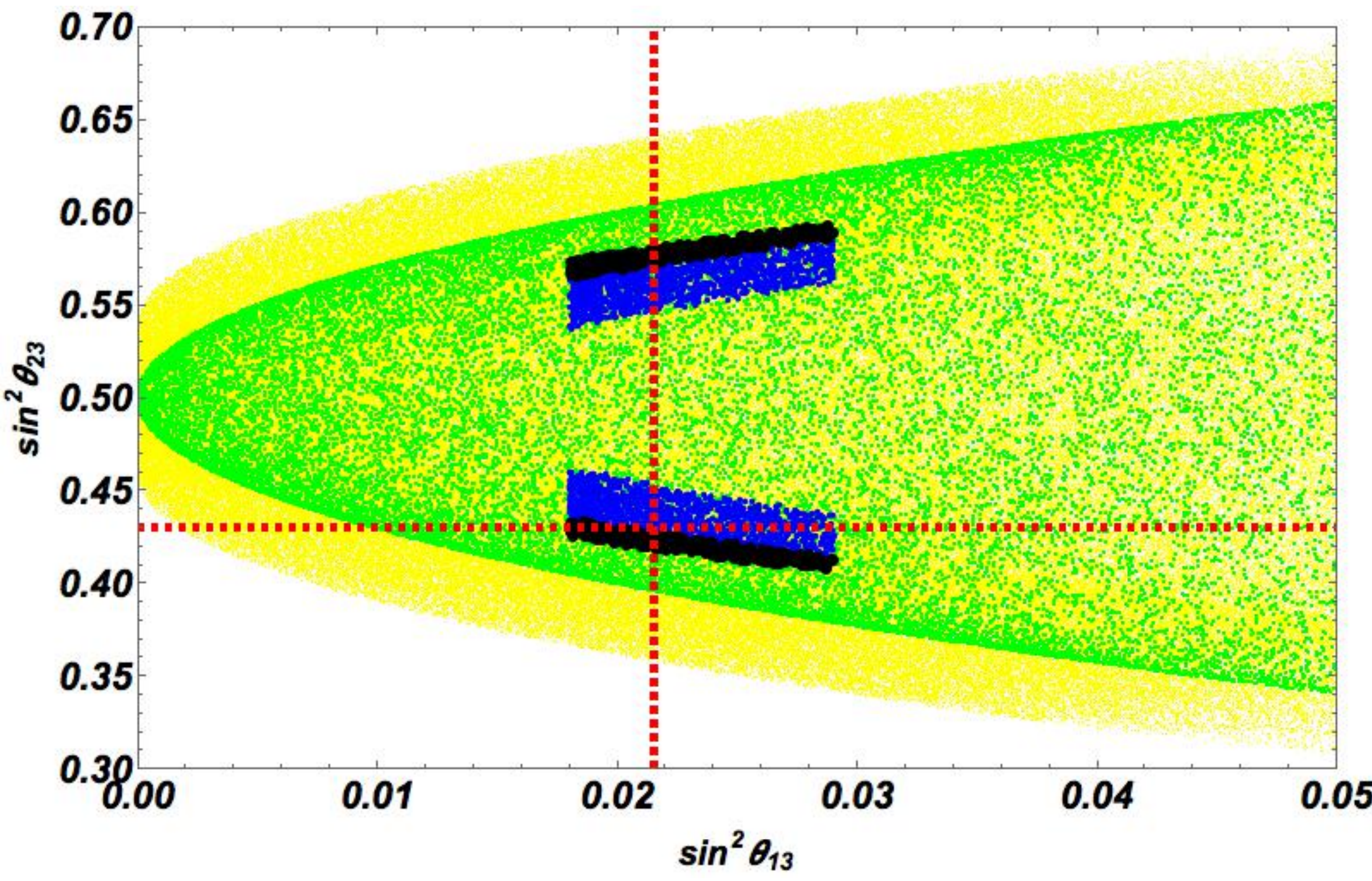}
		\vspace{-60pt}
		\caption{$\sin^2 \theta_{13}$ vs. $ \sin^2 \theta_{23}$. The yellow (green) area represent all values of the calculation (with $Z_{32}=0$), while in blue (black) we have the allowed values after applying all  experimental constrains from Table \ref{tab:summary} (with $Z_{32}=0$). The red grid-lines represent the central values.}
		\label{fig:plot1}
	\end{figure}

\begin{figure}[h]
		\centering
		\includegraphics[width=8cm,height=12cm,angle=270,origin=c]{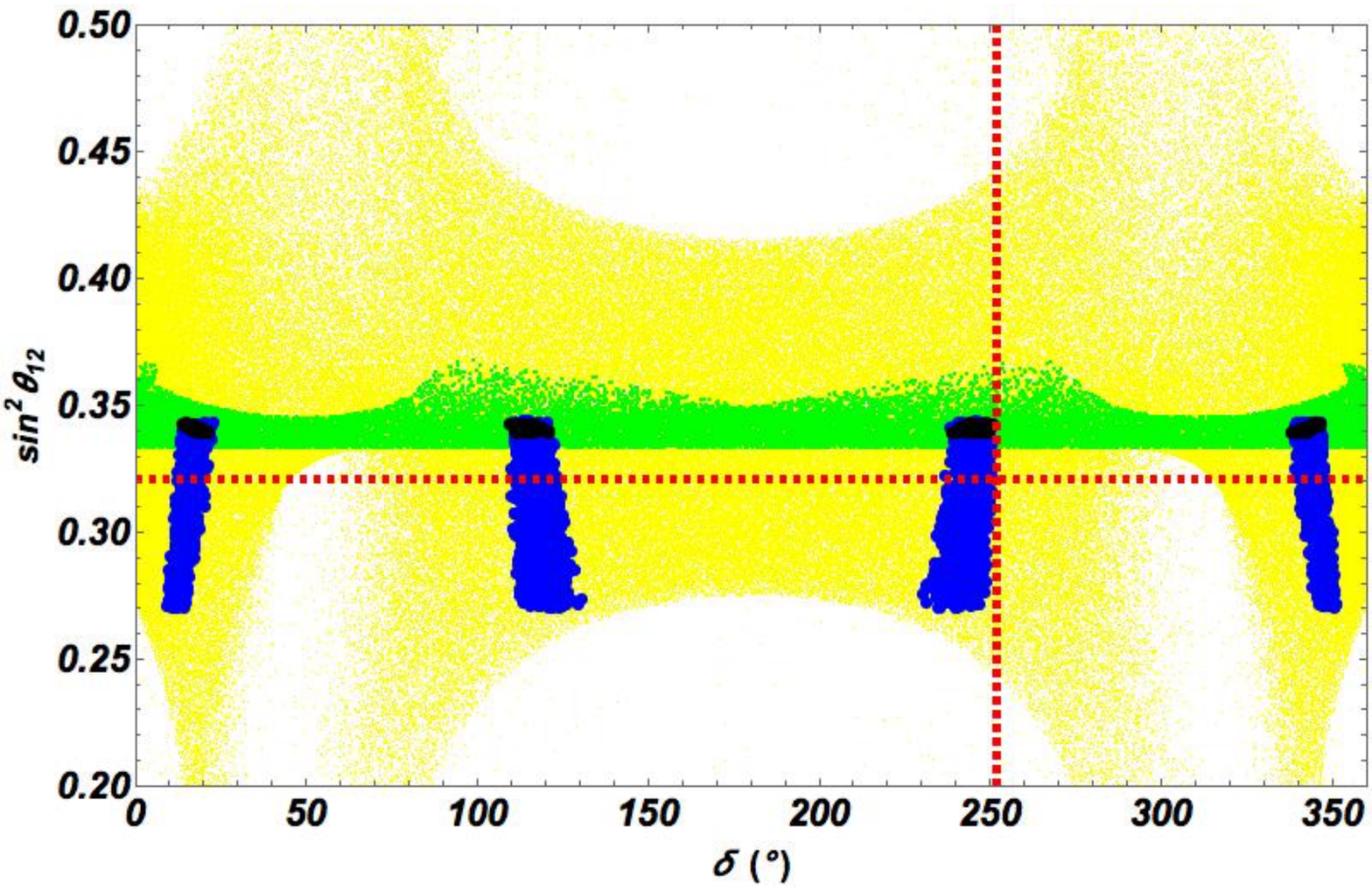}
		\vspace{-70pt}
		\caption{$\sin^2 \theta_{12}$ vs $ \delta$. The color code is the same as in Fig.$\,$\ref{fig:plot1}.}
		\label{fig:plot2}
	\end{figure}

In Figs.$\,$\ref{fig:plot1} and \ref{fig:plot2} we show results using the following fixed values for the parameters: $\mu_{eff} = 1*10^9$, $(\Lambda_5)_{eff}= 2*10^{-3}$, $\Delta_{21}^2 = 7.5*10^{-5}$, $\Delta_{32}^2 = 2.55*10^{-3}$,  $m_{f2} = 5*10^{11}$,  $m_{s2} = 3*10^{14}$ and  $Re(K_1)=Re(K_1)=Re(K_1)=0.1$, while using the following range for the remaining ones: $-0.15<Re(Z_{31}),Im(Z_{31}),Im(K_1)<0.15$. Although these are enough to get all experimental quantities inside 3$\sigma$, we can see from Fig.$\,$\ref{fig:plot2} that we require  at least one more non zero parameter to achieve $sin^2_{12}<1/3$, for this we will use $-0.015<Re(Z_{32})<0.015$. All parameters not mention where set to zero.

	We are using a very reduced region of the parameter space. Specifically, we are setting most of the components of the 2-L Yukawas $Z_{ij}$ to zero, and most of the parameters to be real. We used $Z_{13}$ and $K_1$ as complex numbers; this will allow to get Dirac phases. It can be seen in Fig.$\,$\ref{fig:plot2} that although we can reproduce the central value for this phase, not all values are available, if one wishes to do so, we expect that using cobimaximal mixing \cite{Ma:2015pma}  for the 1-L contribution would be a much better option than tribimaximal.
	
	The scope of this example is not to show all possible parameter space, but to show that the method work, even with a small region being used. 
	One should note that with this region of  parameter space we achieve normal hierarchy for the masses. Nevertheless, given that all experimental values are archived with many parameters set to zero, we expect that the parameter space is much bigger than the one used, and inverted hierarchy can also be achieved.
	
	\section{Method}
	
	Now that we have seen one example of the method and its benefits, we will look at the method as a whole.
	
	In the most generic case the 1-L and 2-L contributions are given by 
	\begin{eqnarray}
	m_{1-L}&=&Y \Lambda_1 Y^T,  \label{masses1L}  \\
	m_{2-L}&=& X_1  \Lambda_{2} X_2^T + X_2  \Lambda_{2}^T X_1^T, \label{masses2L}
	\end{eqnarray}

	where $\Lambda_1$ is the 1-L function and $\Lambda_{2}=\Lambda_{21} X_3 \Lambda_{22}$ is the generalized 2-L function composed of two integrals times a Yukawa. 
	
	We want to remark the fact that $\Lambda_1$ is a diagonal matrix while $\Lambda_2$ is not. This property of the 1-L and 2-L contributions is the reason one can easily perturb and break the texture in 1-L.
	
	Let us assume that we have a neutrino mass matrix  that is composed of two contributions, where $M_{\nu}=m_1+m_2$, is diagonalized in the flavor base by the mixing matrix $U_{\nu}$, $U^T_{\nu}M_{\nu}U_{\nu}=M_{diag}$. Also,  $m_1$  is diagonalized by an exact texture $U_0$, such that $U^T_{0}m_1U_{0}=m_{diag}$, where $U_{\nu}=U_0 \cdot O$ and  $U_0$ can be bimaximal, tribimaximal, cobimaximal, etc.  
	
	Then, using Eq. (\ref{Casas}), one can approximate the perturbation matrix O as 
	\begin{eqnarray}
	O&\simeq& m_d^{1/2} R_1^T  R_2 M_d^{-1/2}  + \frac{1}{2} m_d^{1/2} R_1^T \Lambda_2\Lambda_1^{-1}R_2 M_d^{-1/2} .
	\end{eqnarray}
	
	The first term is close to unity, the second term and therefore the perturbation is proportional to $\Lambda_{2}/ \Lambda_{1}$.

	We can envision three different alternatives of possible models: \textit{i}- The 2-L diagrams can be constructed with the addition of only one scalar field compared to the 1-L. In this case $\Lambda_2$ is symmetric and therefore the neutrino mass is: $M_{\nu}= Y \Lambda_1 Y^T +  Y \Lambda_2 Y^T$. \textit{ii}: The 1-L and 2-L diagrams share no fields in common. The neutrino mass is given by: $M_{\nu}= Y \Lambda_1 Y^T +  X_1  \Lambda_{2} X_2^T + X_2  \Lambda_{2}^T X_1^T $. \textit{iii}- More than one field is required to generate the 2-L from the 1-L diagram, but they do have some fields in common. The neutrino mass has the form: $M_{\nu}= Y \Lambda_1 Y^T +  Y  \Lambda_{2} X_2^T + X_2  \Lambda_{2}^T Y^T $.

	\textit{A priori} all three categories can be used and there is no real physical reason to choose one over another, other than the number of parameters.
	
	%The 1-L and 2-L integrals are:
	%\begin{eqnarray}
	%& &\Lambda_1= \int \frac{d^4k }{(2\pi)^4 \, i} \frac{M_A}{(k^2 - %M_A^2)(k^2 - M_B^2)} \label{eq:lambda1} \\
	%& &\Lambda_{2a} = \int \frac{d^4k }{(2\pi)^4} \int \frac{d^4q %}{(2\pi)^4} \label{eq:lambda2a} \\ & & \frac{8M_A M_C}{(k^2 - %M_A^2)(k^2 - M_B^2)(q^2 - M_C^2)(q^2 - M_D^2)((k-q)^2 - %M_E^2)}  \nonumber \\
	%& &\Lambda_{2b} = \int \frac{d^4k }{(2\pi)^4} \int \frac{d^4q %}{(2\pi)^4} \label{eq:lambda2b} \\ & & \frac{4M_A M_C M_E}{(k^2 - %M_A^2)(k^2 - M_B^2)(q^2 - M_C^2)(q^2 - M_D^2)((k-q)^2 - %M_E^2)} \nonumber
	%\end{eqnarray}
	
	%Where $M_{A,C,E}$ are fermion masses.
	
	A benefit of this method is that the CP violation in the neutrino sector will come directly from the complexity of the Yukawas. Therefore one can reproduce any desired value for the Dirac CP phase. Even if the Yukawas involved in the 1-L diagram are complex. When $\theta_{13}=0$ there will be no CP violation at the 1-L level. But  with the 2-L contribution, the CP phase will start with an initial non-zero value that will be perturbed. If the extra Yukawas used for 2-Ls are complex, then the CP phase will also get perturbed from an initial value set by the 1-L complex Yukawas.

	\section{Conclusions}
	In this letter we have used a new theoretical method for which one can obtain the correct experimental values for the neutrino mixing matrix by perturbing  a texture given from an exact symmetry.
	
	We have shown this method with an example, in which we introduced a model  that reproduces tribimaximal mixing at a one-loop level. With the addition of only one scalar field, we were able to construct a two-loop contribution that is naturally suppressed compared to the original mass, and show that we are able to obtain the neutrino mixing angles and mass differences within a $3\sigma$ deviation (including large Dirac phases).
	Even more models used in this method will contain a possible DM candidate.
	
	In this work I want to encourage the idea that perturbations to neutrinos can have a physical explanation outside of being part of the model itself, or arising from a broken symmetries.
	
	\hspace{.1cm}
	
	{\it Acknowledgments} --
	I would like to  thank Atri Bhattacharya, Julian Heeck, and Ernest Ma for interesting discussions and comments.
	This work was supported by the ``Fonds de la Recherche Scientifique-FNRS'' under 
	grant number 4.4501.15.

\end{document}